\documentclass[11pt]{article}
\usepackage{Blois,graphicx}

\def\Journal#1#2#3#4{{#1} {\bf #2}, #3 (#4)}

\def\EPJC{{\em Eur.\ Phys. J.} C}
\def\NPB{{\em Nucl.\ Phys.} B}
\def\PLB{{\em Phys.\ Lett.}  B}

\def\PRD{{\em Phys.\ Rev.} D}
\def\ZPC{{\em Z. Phys.} C}

\newcommand{\bm}[1]{\mbox{\boldmath $#1$}}

\newcommand{\re}{\mathrm{Re}\,}

\newcommand{\gsim}{\raisebox{-4pt}{$
\,\stackrel{\textstyle >}{\sim}\,$}}

\newcommand{\half}{ {\textstyle\frac{1}{2}} }

\begin{document}

\addtolength{\footskip}{1em}
\pagestyle{plain}

\begin{flushright} \textrm{hep-ph/0509107 \\ DESY-05-158}
\end{flushright}

\vspace*{1.1cm}  
%
\title{AZIMUTHAL DEPENDENCE IN DIFFRACTIVE PROCESSES
\footnote{To appear in the Procs.\ of the XIth International Conference
     on Elastic and Diffractive Scattering, Ch\^{a}teau de Blois,
     France, May 15--20, 2005}
}

\author{M. DIEHL}

\address{Deutsches Elektronen-Synchroton DESY, 22603 Hamburg, Germany}

\maketitle\abstracts{ Azimuthal distributions in high-energy processes
give information about the helicity structure of diffractive
reactions.  I discuss predictions of several dynamical mechanisms in
this context, both for electron-proton and for hadron-hadron
collisions.  }


\section{Azimuthal distributions in $ep$ diffraction}

\begin{figure}[hb]
\begin{center}
\includegraphics[width=0.35\textwidth,bb=0 -40 356 227]{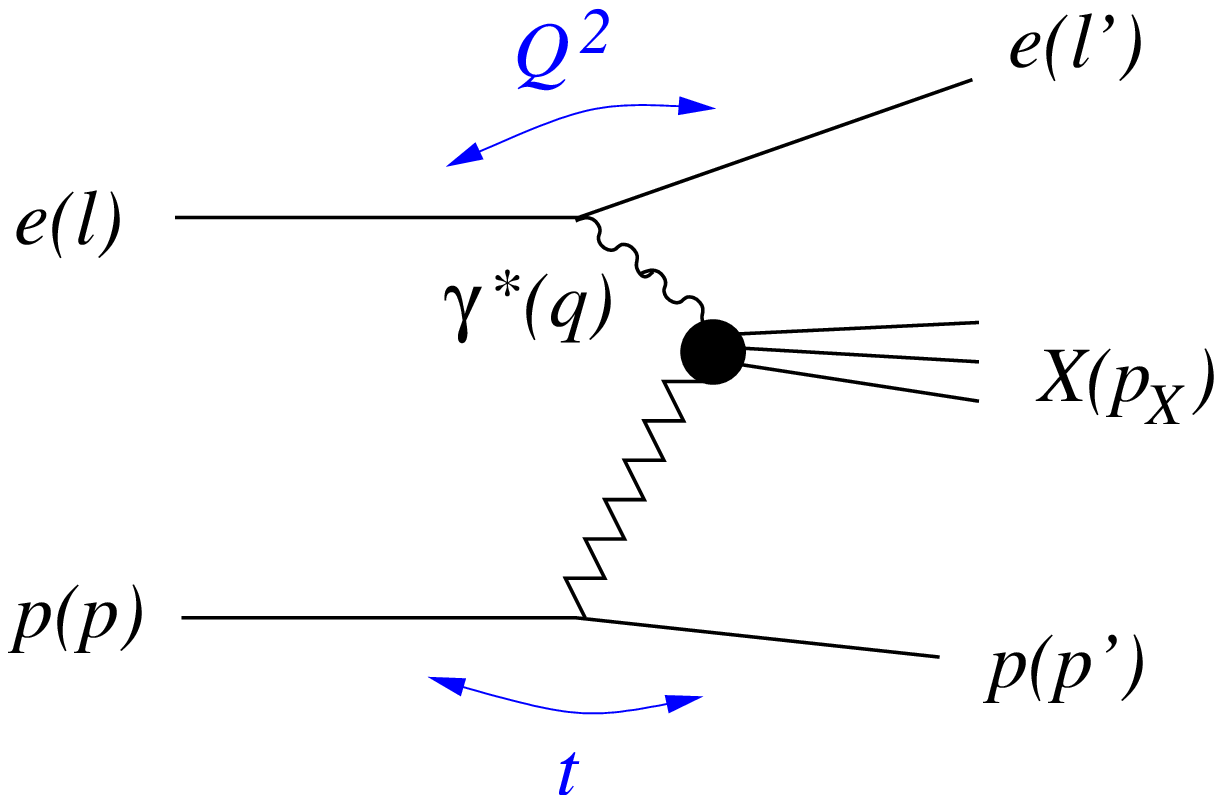}
\hspace{0.02\textwidth}
\includegraphics[width=0.6\textwidth]{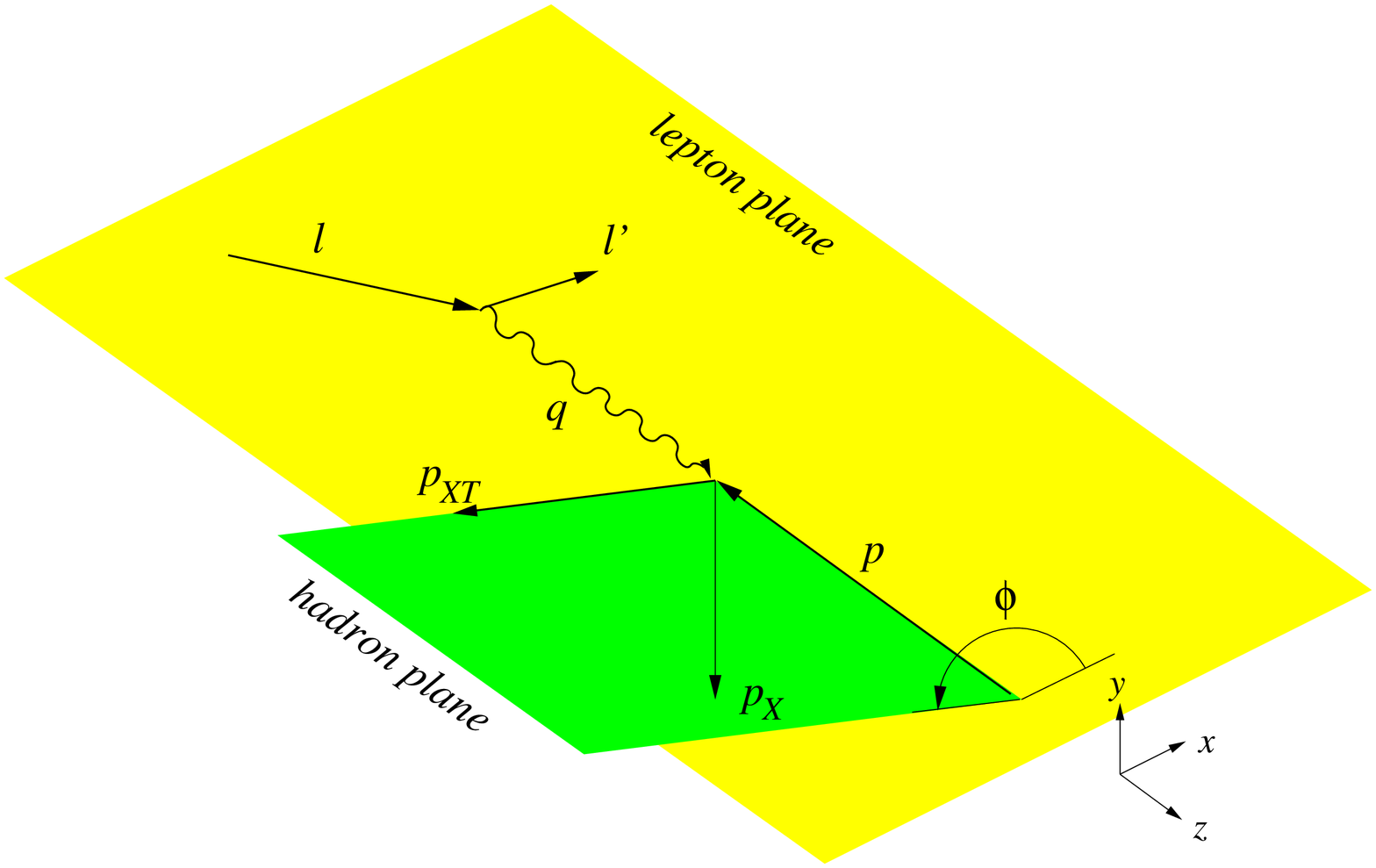}
\caption{\label{fig:ddis-kin} Kinematics for diffractive dissociation
  of a $\gamma^*$ into an inclusive hadronic system $X$ in the
  $\gamma^* p$ c.m.}
\end{center}
\end{figure}

In this talk I give several examples of what azimuthal distributions
can tell us about diffractive interactions.  As a first example let us
look at inclusive diffraction in deep inelastic scattering, $ep\to
eXp$, and consider this process in the $\gamma^* p$ c.m.  We are
interested in the angle $\phi$ between the plane spanned by the hadron
momenta $\bm{p}_X$ and $\bm{p}$ and the one spanned by the lepton
momenta $\bm{l}$ and $\bm{l}'$, as shown in Fig.~\ref{fig:ddis-kin}.

$\phi$ is an angle around the momentum of the virtual photon, and
given the general relation between rotations and angular momentum, it
is not surprising that the dependence of the $ep$ cross section on
$\phi$ contains information on the helicity of the exchanged photon.
Indeed one can write\,\cite{Arens:1996xw}
\begin{equation}
  \label{ep-xsect}
  \frac{d\sigma(ep\to eXp)}{d\phi\, dQ^2\, dy} \propto
  \sigma_{++} + \varepsilon\sigma_{00} 
  - 2\sqrt{\varepsilon (1+\varepsilon)}\, \cos\phi\, \re\sigma_{+0}
  - \varepsilon\, \cos(2\phi)\, \sigma_{+-} \: ,
\end{equation}
where $Q^2= -q^2$ is the photon virtuality, $y= (qp) /(lp)$ the usual
inelasticity parameter, and $\epsilon= (1-y) /(1-y+\half y^2)$ the
ratio of longitudinal and transverse photon flux.  The $\phi$
dependence of the $ep$ cross section is determined by the cross
sections or interference terms $\sigma_{ij}$ for the hadronic
subprocess $\gamma^* p\to Xp$ with specified photon helicities,
\begin{equation}
\sigma_{ij} \propto {\sum\nolimits_{X}}\,
  {\sum\nolimits_{\mathrm{spins}}}\;
  \mathcal{A}_i^* \mathcal{A}_j^{\phantom{*}} ,
\end{equation}
where $\mathcal{A}_i$ is the amplitude for $\gamma^* p\to Xp$ with
photon helicity $i$, and the sums run over final states $X$ and the
spins of initial and final proton.  The $\sigma_{ij}$ only
depend on the kinematics of the $\gamma^* p$ subprocess but not on
$\phi$.  Both sides of (\ref{ep-xsect}) can be made differential in
further variables describing the hadronic final state, such as the
invariant momentum transfer $t= (p-p')^2$ to the proton or the
invariant mass $M_X$ of the inclusive system $X$.

{}From the Schwarz inequality $|\sigma_{ij}|^2 \le \sigma_{ii}\,
\sigma_{jj}$ it is clear that the interference terms in
(\ref{ep-xsect}) are bounded by the transverse and longitudinal cross
sections $\sigma_{++}=\sigma_T$ and $\sigma_{00}=\sigma_L$.  More
stringent bounds are obtained by using that $\sigma_{ij}$ is a
positive semidefinite $3\times 3$ matrix, since for arbitrary
coefficients $c_i$ the linear superposition $\sum_{ij} c_i^*\,
\sigma_{ij}^{\phantom{*}}\, c_j^{\phantom{*}}$ is a cross section and
hence cannot be negative.  The positivity condition can be written
as\,\cite{Arens:1996xw,Diehl:1997jp}
\begin{eqnarray}
&& \varepsilon\sigma_L \:\le\: \sigma_\varepsilon + \sigma_{+-} \: ,
\nonumber \\[0.3em]
&&
\varepsilon\sigma_L \:\le\: 
\half (\sigma_\varepsilon - \sigma_{+-}) +
\half \sqrt{ (\sigma_\varepsilon - \sigma_{+-})^2
       - 8\varepsilon\, (\re \sigma_{+0})^2 } \: ,
\nonumber \\
&&
\varepsilon\sigma_L \:\ge\: 
\half (\sigma_\varepsilon - \sigma_{+-}) -
\half \sqrt{ (\sigma_\varepsilon - \sigma_{+-})^2
       - 8\varepsilon\, (\re \sigma_{+0})^2 } \: ,
\hspace{3em}
\end{eqnarray}
where $\sigma_{\varepsilon} =\sigma_T + \varepsilon\sigma_L$.  The
right-hand sides of these inequalities can directly be extracted from
the $\phi$ dependence of the $ep$ cross section and provide upper or
lower bounds on the longitudinal cross section $\sigma_L$, whose
direct extraction from $\sigma_{\varepsilon}$ requires a Rosenbluth
separation and thus measurements at different $ep$ collision energies.
The interference terms must be large in size for these bounds to be
useful.  If they are however small, one cannot draw strong
conclusions: two amplitudes may both be large but not interfere
because of their relative phase, or there can be cancellations between
positive and negative interference in the sum over final states.

\begin{figure}
\begin{center}
\includegraphics[width=0.82\textwidth]{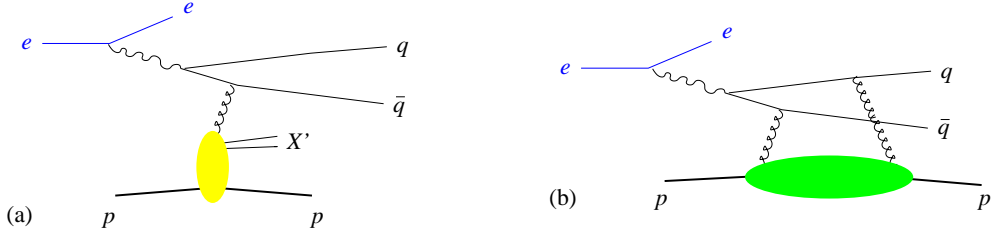}
\caption{\label{fig:ddis-graphs} (a) Graph for diffractive scattering
   in the collinear factorization framework.  The blob symbolizes the
   diffractive gluon density.  (b) Graph for diffraction proceeding
   through hard two-gluon exchange.  The blob represents the
   generalized gluon distribution of the proton.}
\end{center}
\end{figure}

The $\phi$ dependence of the diffractive cross section has been
measured by ZEUS, and no $\cos\phi$ or $\cos(2\phi)$ modulation was
found within experimental errors.\,\cite{Chekanov:2004hy} What does
theory predict for the $\phi$ dependence?  For inclusive diffraction
there is a factorization theorem,\,\cite{Collins:1997sr} valid in the
Bjorken limit of large $Q^2$ at fixed $\beta= Q^2/(M_X^2+Q^2)$, $x_B=
Q^2/(2pq)$ and $t$.  In this limit the $ep$ cross section can be
calculated as a convolution of diffractive parton densities in the
proton with the cross section for electron scattering on the
corresponding parton, see Fig.~\ref{fig:ddis-graphs}a.  Essential in
our context is that the parton-level cross section (for $eg\to e
q\bar{q}$ in the example of the figure) is evaluated with the
transverse momentum of the incoming parton approximated by {zero} in
the $\gamma^* p$ c.m.  As a consequence, the parton-level process
receives no information on the outgoing proton momentum, and the
resulting cross section cannot depend on $\phi$, for whose definition
this momentum is essential.  In the Bjorken limit, the $\phi$
dependence of the $ep$ cross section is thus indeed predicted to be
flat.  On the other hand, both the longitudinal and the transverse
diffractive structure functions $F_L^D$ and $F_T^D$ (related to the
$\gamma^* p$ cross sections by a kinematical factor) are nonzero and
become $Q^2$ independent in that limit, up to logarithmic corrections.

A more complicated picture is obtained when one considers diffractive
production of a $q\bar{q}$ pair by two-gluon exchange, shown in
Fig.~\ref{fig:ddis-graphs}b.  Corresponding calculations provide a
good description of inclusive diffraction for $\beta\gsim 0.5$,
whereas for small $\beta$ diffractive final states with additional
gluons become important.\,\cite{Wusthoff:1999cr} (We note that such
calculations are sensitive to infrared physics when the produced quark
has small transverse momentum, but shall not dwell on this point
here.)  The two-gluon exchange mechanism gives a longitudinal
structure function $F_L^D$ which at given $\beta$ falls like $1/Q^2$
but contrary to $F_T^D$ does not vanish for $\beta\to 1$.  At large
$\beta$, this twist-four contribution to the $ep$ cross section is
hence potentially dangerous for analyses based on the twist-two
factorization theorem discussed above, and information on the
importance of $F_L^D$ in given kinematics is highly important.  Such
information may be provided by the $\phi$ distribution: calculations
of two-gluon exchange predict an interference structure function
$F_{+0}^D$ which is suppressed by $\sqrt{-t} /Q$ compared with $F_T^D$
but remains finite for $\beta\to
1$.\,\cite{Diehl:1996st,Nikolaev:1998wf} The corresponding $\cos\phi$
modulation of the cross section at large $\beta$ (\textit{i.e.}\ for
$Q^2 \gg M_X^2$) is not seen in the ZEUS data,\,\cite{Chekanov:2004hy}
where the bin with the highest $\beta$ is centered around $0.73$.  We
remark however that in $ep\to e\rho p$ at large $Q^2$, where the
inclusive hadron system $X$ is replaced by a single $\rho$ meson, a
significant $\cos\phi$ modulation has indeed been measured and is well
described by calculations based on the two-gluon exchange
mechanism.\,\cite{Breitweg:1999fm,Adloff:1999kg}

Let us now turn to the case where the diffractive final state contains
a pair of high-$p_T$ jets.  The jet momenta in the $\gamma^* p$ c.m.\
define a hadron plane different from the one in
Fig.~\ref{fig:ddis-kin}.  The dependence of the cross section on the
azimuth $\phi^{\,\mathit{jj}}$ between this new plane and the lepton
plane is described by an expression of the form (\ref{ep-xsect}), with
appropriate $\gamma^* p$ cross sections and interference terms.  We
can distinguish two types of final states:
\begin{enumerate}
\item \emph{inclusive} dijet production, $ep\to ep +\mathit{jet}
  +\mathit{jet} +X'$, with a inclusive system $X'$ of hadrons in the
  direction of the initial proton.  This can be described in the same
  diffractive factorization formalism as the inclusive cross section
  (see Fig.~\ref{fig:ddis-graphs}a).  Independent of the diffractive
  quark and gluon densities, this mechanism gives a negative
  interference term\,\cite{Bartels:1996tc,Diehl:1997jp}
  \begin{equation}
    \sigma_{+-}^{\,\mathit{jj}} = -
    \half\sigma_{L}^{\,\mathit{jj}}
  \end{equation}
  and thus predicts a $\cos(2\phi^{\,\mathit{jj}})$ modulation such
  that the dijets are preferentially produced \emph{in} the lepton
  plane.  In kinematics where diffractive factorization is valid, this
  modulation allows one to extract the longitudinal cross section
  without Rosenbluth separation, and thus provides extra constraints
  on the diffractive parton densities.
\item \emph{exclusive} dijet production, $ep\to ep +\mathit{jet}
  +\mathit{jet}$.  Such events are expected to become important for
  $Q^2\gg M_X^2$ but have not yet been isolated experimentally.  The
  two-gluon exchange mechanism of Fig.~\ref{fig:ddis-graphs}b gives a
  positive interference term\,\cite{Bartels:1996tc,Braun:2005rg}
  \begin{equation}
    \sigma_{+-}^{\,\mathit{jj}} = 
    \frac{2r}{1 - 2r}\, \sigma_T^{\,\mathit{jj}} ,
    \qquad\qquad\qquad
    r=\frac{p_T^2}{M_X^2} ,
  \end{equation}
  where $p_T$ is the transverse jet momentum in the $\gamma^* p$ c.m.
  This mechanism preferentially produces jets \emph{perpendicular to}
  the electron plane.
\end{enumerate}
To distinguish the two types of final states at hadron level is not
trivial, especially if the system $X'$ is not very energetic.  The
different $\phi^{\,\mathit{jj}}$ distribution of the two production
mechanisms can provide a clear distinction and thus help to establish
which dynamical description is adequate in given kinematics.  We note
that both mechanisms also predict a $\cos\phi^{\,\mathit{jj}}$
modulation of the cross section.


\section{Diffraction in $pp$ collisions}

Diffraction in $pp$ collisions is more complex than in $ep$
collisions, even in the presence of a hard scale.  There are no
factorization theorems like the ones we encountered in the previous
section, because of soft interactions between partons in the two
colliding hadrons.  To describe the dynamics, one presently has to
rely on assumptions or models, and we will see that azimuthal
distributions can be valuable to test and develop these.

\begin{figure}
\begin{center}
\includegraphics[width=0.30\textwidth]{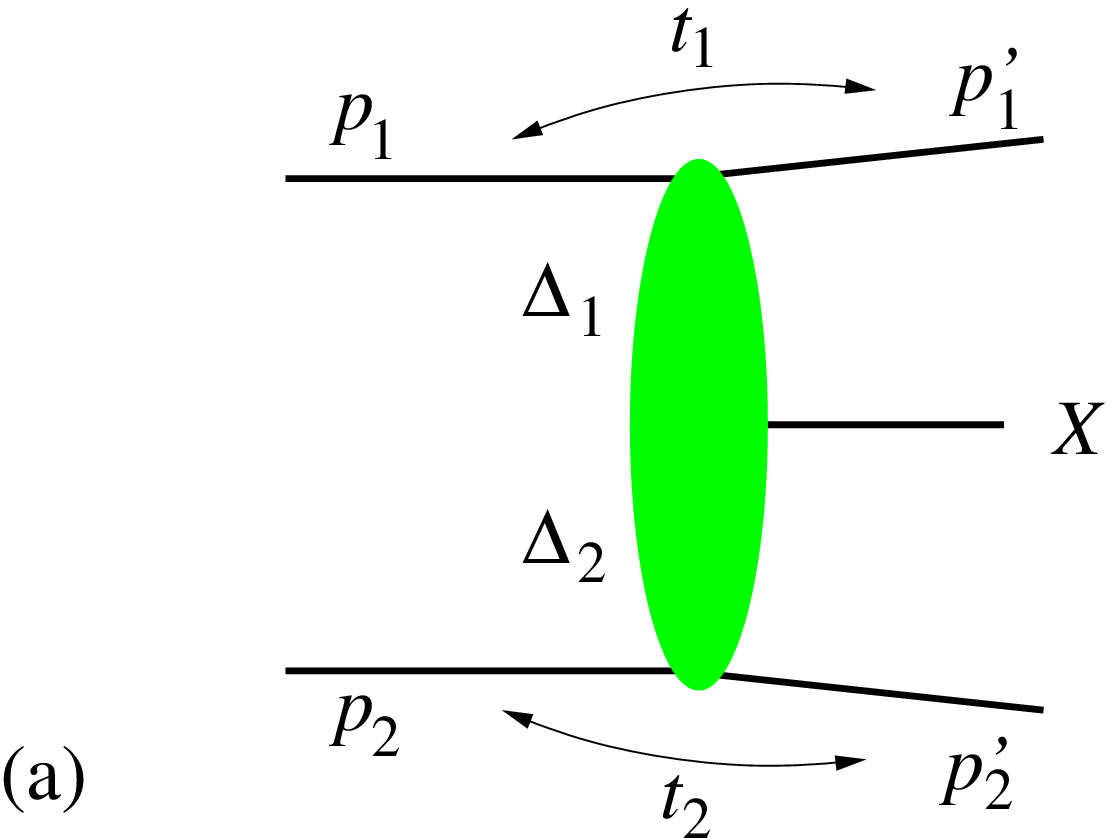}
\hspace{0.1\textwidth}
\includegraphics[width=0.33\textwidth]{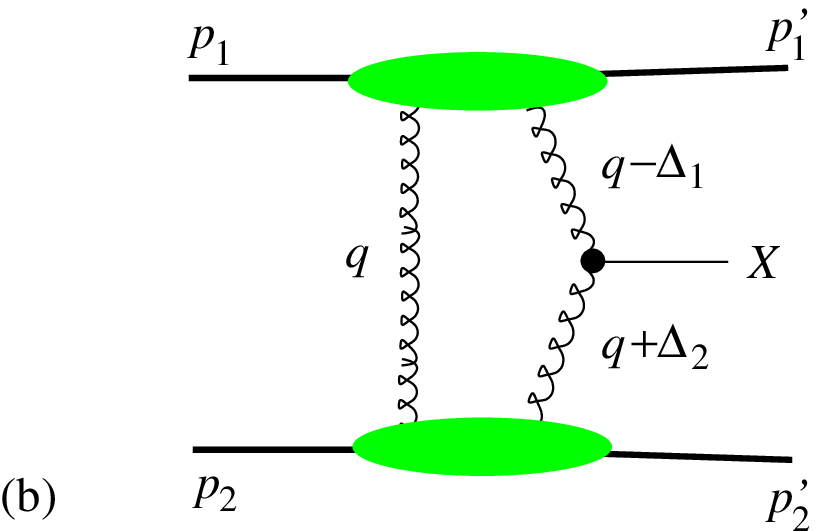}
\caption{\label{fig:diff-pp} (a) Schematic diagram for diffractive
  production of a particle or system of particles $X$ in a $pp$
  collision.  (b) A simple graph for this process in the two-gluon
  exchange mechanism.  $p_i^{\protect\phantom{'}}$, $p_i'$, $\Delta_i$
  and $q$ denote four-momenta.}
\end{center}
\end{figure}

Consider diffractive production of a particle or system of particles
$X$ in the $pp$ c.m.  The azimuthal angle $\phi$ between the plane
spanned by $\bm{p}_1$ and $\bm{p}_1'$ and the one spanned by
$\bm{p}_2$ and $\bm{p}_2'$ (with momenta as shown in
Fig.~\ref{fig:diff-pp}a) contains information on the helicity
transferred in the $t_1$ and $t_2$ channels.\footnote{There are subtle
issues concerning the difference between $\phi$ and the angles between
the $\bm{p}_1$-- $\bm{p}_1'$ and $\bm{p}_2$-- $\bm{p}_2'$ planes in
the rest frame of $X$ or in the c.m.\ of $X$ and $p_2'$.  This
difference is negligible if the invariant momentum transfers $t_1$ and
$t_2$ are much smaller than the squared invariant mass $M_X^2$ of $X$,
which we assume here.\,\protect\cite{Arens:1996xw,Close:1999is}}
As a simple ansatz one may assume a factorized form of the cross
section,
\begin{equation}
  \label{regge-fact}
\sigma(pp\to pXp) = \sum_{i_1 j_1,i_2 j_2}
  \rho_{j_1 i_1}(p_1\to p_1' I\!\!P_1^{\phantom{'}})\, 
  \rho_{j_2 i_2}(p_2\to p_2' I\!\!P_2^{\phantom{'}})\, 
  \sigma_{i_1 j_1,i_2 j_2}(I\!\!P_1^{\phantom{'}} 
                           I\!\!P_2^{\phantom{'}}\to X) ,
\end{equation}
where $j_1$ ($j_2$) and $i_1$ ($i_2$) respectively denote the
helicities transferred in the $t_1$ ($t_2$) channels in the amplitude
and its complex conjugate.  The physical picture behind this is that
each of the colliding protons ``emits'' a pomeron, and that the two
pomerons fuse to produce $X$, as our symbolic notation in
(\ref{regge-fact}) suggests.  $\rho_{j_1 i_1}$ and $\rho_{j_2 i_2}$
play the roles of spin-density matrices of the pomerons, whose fusion
into $X$ is described by cross sections and interference terms
$\sigma_{i_1 j_1,i_2 j_2}$.  If the pomeron behaves like a spin 1
exchange, the helicity indices are restricted to values $0$ and $\pm
1$.  This is for instance the case in the Donnachie-Landshoff pomeron
model.\,\cite{Donnachie:1984xq} Close \textit{et al.} performed a
general analysis for the production of a resonance $X$ with quantum
numbers $J^P$ under these assumptions and found a $\phi$
dependence\,\cite{Close:1999is,Close:1999bi}
\begin{eqnarray}
\sigma(0^-) &\propto &
   t_1 t_2 \sin^2\!\phi\, |\mathcal{A}_{++}|^2 ,
\nonumber \\
\sigma(0^+) &\propto & t_1 t_2\, |\mathcal{A}_{00} 
   + \ldots \mathcal{A}_{++}|^2 ,
\nonumber \\
\sigma(1^+) &\propto &  t_1 t_2 \sin^2\!\phi\, |\mathcal{A}_{++}|^2
 + | \ldots \mathcal{A}_{+0} + \ldots \mathcal{A}_{0+} |^2 ,
\end{eqnarray}
where $\mathcal{A}_{i_1 i_2}$ is the amplitude for two-pomeron fusion
into the resonance, and the dots denote coefficients depending on
$\phi$.  Results for spins $J=2$ and $3$ were also obtained.  These
results agree with a general analysis in Reggeon field
theory\,\cite{Kaidalov:2003fw} but are more restrictive due to the
specific assumptions on the nature of pomeron exchange.  The ratio
$\mathcal{A}_{00} / \mathcal{A}_{++}$ of longitudinal and transverse
amplitudes depends on details of how the pomerons couple to the
produced resonance, and the $\phi$ dependence in $0^+$ and $2^+$
production has been proposed to discriminate glueball from
quark-antiquark bound states.\,\cite{Close:1999is,Close:1999bi}

If the pomeron behaves like a spin 1 exchange, an important question
is whether the vector current describing its coupling to particles is
conserved or not.  If the pomeron couples like a conserved current,
one finds $\mathcal{A}_{0\hspace{0.5pt}i_2} \sim \sqrt{-t_1}$ at small
$t_1$, whereas a behavior $\mathcal{A}_{0\hspace{0.5pt}i_2} \sim
1/\sqrt{-t_1}$ is obtained for a non-conserved current (which is for
instance realized in the Donnachie-Landshoff
model).\,\cite{Arens:1996xw} Measurements of $f_1$ production disfavor
a conserved current, whereas the assumption of a non-conserved current
can accommodate data for various $f_0$, $f_1$, $f_2$, $\eta_2$
resonances.\,\cite{Close:1999bi}

In QCD, pomeron exchange becomes the exchange of a pair of interacting
gluons.  A simple graph for the process $pp\to pXp$ is shown in
Fig.~\ref{fig:diff-pp}b, where the blobs at the top and bottom
describe interactions between the gluons and their coupling to the
proton.  $X$ is now produced by the fusion of two gluons instead of
two pomerons, whereas another gluon is directly exchanged between the
colliding protons.  Hence, factorization as in (\ref{regge-fact}) does
not hold and one has instead
\begin{equation}
  \label{two-glue}
\sigma(pp\to pXp) = \int d^2 q_T^{\phantom{*}}\, d^2 q_T^{*} 
  \sum_{i_1 j_1,i_2 j_2}
  \rho_{j_1 i_1, j_2 i_2}(p_1 p_2 \to p_1' p_2'\,
                            g_1^{\phantom{'}} g_2^{\phantom{'}}) \,
  \sigma_{i_1 j_1, i_2 j_2}(g_1^{\phantom{'}} g_2^{\phantom{'}}\to X) .
\end{equation}
Here $q$ and $q^*$ are loop momenta appearing in the amplitude and its
complex conjugate, respectively, and the subscript $T$ denotes their
transverse components in the $pp$ c.m.  The two gluons producing $X$
couple with a conserved current, due to their nature as gauge
particles, but they do not carry the full momentum $\Delta_1$
($\Delta_2$) exchanged in the $t_1$ ($t_2$) channel.  In particular,
the typical values of their virtualities in the loop integrals
(\ref{two-glue}) remain finite if $t_1$ or $t_2$ goes to zero.  The
mechanism shown in Fig.~\ref{fig:diff-pp}b is thus not in
contradiction with the findings from meson production discussed in the
previous paragraph.

In diffractive production of meson resonances there is no hard scale,
and the graph shown in Fig.~\ref{fig:diff-pp}b is to be interpreted in
the sense of a non-perturbative model, such as for instance the one by
Landshoff and Nachtmann.\,\cite{Landshoff:1986yj} If however $X$ is a
Higgs boson or another heavy particle, the hard scale $M_X$ allows at
least part of the dynamics to be described in perturbation theory.
This has been elaborated by the Durham
group\,\cite{Kaidalov:2003fw,Khoze:2000cy} and was recently reviewed
by Forshaw.\,\cite{Forshaw:2005qp} The blobs in
Fig.~\ref{fig:diff-pp}b then represent the generalized gluon
distribution $f_g$ of the proton, and the corresponding scattering
amplitude has the form\,\cite{Kaidalov:2003fw}
\begin{equation}
  \label{higgs-master}
\mathcal{A}(J^P) \propto
  \int d^2 q_T\, \frac{V(J^P)}{q_T^2\, (q_T-\Delta_{1T})^2 \,
  (q_T+\Delta_{2T})^2} \, 
  f_g(q_T, -\Delta_{1T})\, f_g(q_T, \Delta_{2T}) ,
\end{equation}
where we have omitted the dependence of $f_g$ on the longitudinal
gluon momenta.  Infrared convergence of the integral is ensured by
Sudakov form factors included in $f_g$, which have significant effects
for large $M_X$, but the integral does have some sensitivity to the
infrared region.\,\cite{Forshaw:2005qp} The vertex factor $V(J^P)$
for two gluons coupling to a Higgs depends on its parity,
\begin{equation}
 V(0^+) = (q_T-\Delta_{1T}) \cdot (q_T+\Delta_{2T}) \,,
\qquad\qquad
 V(0^-) = [\, (q_T-\Delta_{1T}) \times (q_T+\Delta_{2T}) \,]_z \,.
\end{equation}
If the transverse momenta $\Delta_{1T}$ and $\Delta_{2T}$ of the
scattered protons are small enough, they can be neglected compared
with $q_T$ and one approximately obtains
\begin{eqnarray}
  \label{higgs-phi}
\mathcal{A}(0^+) &\propto &
  \int \frac{d q_T^2}{q_T^4} 
  f_g(q_T, -\Delta_{1T})\, f_g(q_T, \Delta_{2T}) ,
\nonumber \\
\mathcal{A}(0^-) &\propto & 
  [ \Delta_{1T} \times \Delta_{2T} ]_z \, \int \frac{d q_T^2}{q_T^6}\,
  f_g(q_T, -\Delta_{1T})\, f_g(q_T, \Delta_{2T})
\end{eqnarray}
from (\ref{higgs-master}).  Up to small modulations, the cross section
at small $t_1$ and $t_2$ is hence flat for a scalar Higgs and behaves
like $( \Delta_{1T} \times \Delta_{2T} )^2 \approx t_1 t_2 \sin^2\phi$
for a pseudoscalar one.

A significant modulation of the $\phi$ dependence compared with
(\ref{higgs-phi}) can originate from rescattering of partons in the
colliding protons.  Such interactions are known to have an important
effect on the overall size of the cross section.  The dependence of
the cross section on $\phi$ (as well as on $t_1$ and $t_2$) can be
used to test specific rescattering models, where a crucial aspect
tested is how much transverse momentum compared with $\Delta_{1T}$ and
$\Delta_{2T}$ is transferred by rescattering.  A model study by the
Durham group concluded that even when taking these effects into
account, the $\phi$ dependence still provides a means to discriminate
between different quantum number assignments for newly discovered
particles.\,\cite{Kaidalov:2003fw}


\section{Summary}

We have presented several cases where azimuthal correlations in the
final state can yield valuable insight into diffractive dynamics.  In
$ep$ collisions, the cross section dependence on a suitably defined
angle reflects the helicity of the exchanged virtual photon.
Depending on how significant this dependence is for particular final
states and kinematics, it can provide useful bounds on the cross
section for longitudinal photons, without the need to measure at
different $ep$ collision energies.  It may give an indication for the
importance of higher-twist contributions to the longitudinal
diffractive structure function.  In diffractive dijet production, the
sign of a $\cos(2\phi^{\,\mathit{jj}})$ modulation can distinguish
between exclusive and inclusive production mechanisms.

Azimuthal correlations between the outgoing protons in exclusive
diffractive $pp$ collisions may provide a valuable tool to determine
the parity of new particles such as the Higgs.  Analysis of data on
diffractive production of meson resonances under the assumption of
factorization as in (\ref{regge-fact}) indicates that the predominant
helicities transferred by diffractive exchange are $0$ and $\pm 1$,
and that the current describing how the exchange couples to particles
is not conserved.  A simple mechanism for $pp\to pXp$ in a microscopic
description is two-gluon exchange, where one gluon does not
participate in the production of the particle $X$.  Additional
rescattering between the colliding systems influences the azimuthal
distribution of the final-state protons.  This provides a handle to
validate assumptions and models for these predominantly soft
interactions, which play a major role in diffractive hadron-hadron
scattering.


\end{document}